\documentclass{aa}
\usepackage{txfonts}
\usepackage{graphicx}

\begin{document}

\title{The first wide ultracool binary dwarf in the field: 
DENIS-J055146.0$-$443412.2 (M8.5 $+$ L0) 
\thanks{Based on observations collected at the European Southern
Observatory, La Silla, Chile.}} 

\subtitle{}

\author{M. Bill\`eres \inst{1} \and X. Delfosse \inst{2} \and J.-L Beuzit
\inst{2} \and T. Forveille \inst{3,2} \and L. Marchal\inst{2}
\and E.L. Mart\'{\i}n\inst{4}}

\institute{European Southern Observatory, Casilla 19001, Santiago 19, Chile
\and 
               Laboratoire d'Astrophysique de Grenoble,
               Observatoire de Grenoble,
               BP53,
               F-38041 Grenoble,
               France
\and
               Canada-France-Hawaii Telescope Corporation,
               65-1238 Mamaloha Highway,
               Kamuela, HI 96743,
               U.S.A.
\and
               Instituto de Astrof\'{\i}sica de Canarias, 
               38200 La Laguna, Spain
}

\date{Received / Accepted}

\abstract{We present observations of a new very low mass field binary, 
discovered during an infrared imaging survey of 250 DENIS L and very late-M 
dwarfs. DENIS-J055146.0$-$443412.2 is an M8.5~$+$~L0 pair, with a physical 
separation of over 200~AU. This makes it the widest very low mass binary 
known in the field, by an order of magnitude. Such a system is fragile,
and it would not have survived a close encounter with a third body. Its
existence demonstrate that some very low mass stars/brown dwarfs form 
without ejection from a multiple system, or any other strong dynamical 
interaction. 
}

\authorrunning{Bill\`eres et al.}
\maketitle

\section{Introduction}

Stellar multiplicity is a key observational diagnostic of the star 
formation process, and of the early dynamical evolution
of stellar systems. Different formation scenarios produce distinct
multiple star fractions, different distribution functions for
separations and mass ratios, and distinct variations of these
quantities with primary mass.

The multiplicity of ultracool dwarfs (late-M, L and T dwarfs) has
recently been of particular interest, in part because the leading
scenario for their formation makes a very specific prediction. In
this ``embryo-ejection'' scenario (Reipurth \& Clark \cite{reip01}), 
dynamical interaction in small sub-clusters eject stellar embryo from 
their gas reservoir before they can accrete enough mass to become more 
massive stars. The only binary brown dwarfs expected in this scenario
are close binaries that have enough binding energy to survive the 
N-body interactions, and in SPH simulations (e.g. Bate et al. 
\cite{bate03}) the few BD binaries produced do have separations 
below 10~AU.

High resolution imaging surveys (Mart\'{\i}n, Brandner \& Basri 
\cite{martin99}; Close et al. \cite{close03}; Bouy et
al. \cite{bouy03}; Gizis et al. \cite{gizis03}; Burgasser et
al. \cite{burg03}; Siegler et al. \cite{sieg05}) have found a
few dozen field ultracool binaries, and indicate that $\sim$10-15\%
of the ultracool field dwarfs have a visual companion beyond 
$\sim$~1~AU. None of these binaries is wider than 20 AU, however,
in sharp contrast with observed separations above 1000~AU for
some systems of early/mid-M dwarfs (Marchal et al. \cite{march03}, and 
in prep.). Imaging of brown dwarfs in young clusters (Neuh\"auser et al. 
\cite{Neuh02}; Mart\'{\i}n et
al. \cite{martin03}; Luhman, McLeod \& Goldenson \cite{luh2005}; 
Lucas, Roche \& Tamura \cite{lucas05}) demonstrate
that wide binaries are already rare at this early stage. The lack of 
wide very low mass binaries has generally been considered a vindication
of the ejection scenario, but the observational statistics are still
limited and they do not exclude a low-level tail of wide binaries.

Very recently, Luhman (\cite{luh2004}) indeed reported the discovery of a 
240 AU binary brown dwarf in the Chameleon I star forming region,
and Chauvin et al. (\cite{chau04}, \cite{chau05}) described 
a planetary mass object orbiting 55 AU from a young brown 
dwarf in the TW Hydrae Association. The two systems are dynamically fragile,
and the strong interactions during ejection from a small
sub-cluster would have disrupted them very easily. Both are thus 
strong evidence against all brown dwarfs forming through ejection.
Some skepticism has occasionally been expressed on the physical 
association of the two Luhman brown dwarfs (e.g. Burgasser et al. 
\cite{burg05}), but in our view the case is fairly solid. 
Both systems were on the other hand found somewhat serendipitously, 
making the fraction of brown dwarfs with wide companions uncertain.
Better ascertaining this small fraction needs homogenous observations
of a large sample of ultracool dwarfs.

We are conducting such an observing program for $\sim$250~dwarfs, and 
present here its first result, the discovery of the third wide ultracool 
dwarf binary, and the first in the field. Sec.~2 briefly presents our 
overall imaging program, which will be discussed more fully in a later 
publication, as well as its data for DENIS-J055146.0$-$443412.2 
(hereafter DENIS-J055$-$44). Sec.~3 presents our infrared spectroscopy of the 
new binary, and Sec.~4 discusses the implications of its discovery 
for our understanding of stellar formation.

\begin{table*}
\begin{center}
\begin{tabular}{lllllll} \hline
\multicolumn{7}{c}{Astrometry and photometry for the combined DENIS-J055$-$44
 binary } \\
$\alpha$ (J2000) & $\delta$ (J2000) &  I$_{\rm DENIS}$ & J$_{\rm DENIS}$ & J$_{\rm 2MASS}$ & H$_{\rm 2MASS}$ & K$_{\rm 2MASS}$ \\ 
 05:51:46.0      & -44:34:12.6      &  18.4$\pm$0.3 & 15.5$\pm$0.2 & 15.8$\pm$0.1 & 15.2$\pm$0.1 & 14.9$\pm$0.1 \\ \hline
\multicolumn{7}{c}{NTT measurements : observational elements of the binary} \\
 & $\rho ('')$ & $\theta$ (deg) & $\Delta$J & $\Delta$K & & \\ 
 & 2.20$\pm$0.05 & 358.6 & 0.55$\pm$0.02 & 0.53$\pm$0.02 & & \\ \hline
\end{tabular}
\end{center}
\caption{Photometry and position of the DENIS-J055146.0$-$443412 system. $I$,
$J$, $H$ and $K$ 
are for the combined binary system from the DENIS and 2MASS catalogs;
$\Delta J$, $\Delta K$ and the relative positions were obtained from the 
SOFI images using the point source mode of the MISTRAL myopic deconvolution 
package (Mugnier, Fusco \& Conan \cite{mugnier04}).
}
\label{table_phot}
\end{table*}

\section{NTT imaging search for wide companions of ultracool dwarfs}

The imaging sample is drawn from late-M and L dwarfs identified by
the DENIS survey, which demonstrated its effectiveness at selecting
such objects with the discovery of the first field brown dwarfs
(Delfosse et al. \cite{delf97}). Delfosse et al. (\cite{delf03})
assembled a sample of 300 objects with $I-J>3.0$ (M8 and cooler),
from 5700 squares degrees of DENIS survey data. To date we have 
obtained moderately deep infrared images in the $J$ and 
$K$ bands for 250 of those 300 DENIS ultracool dwarfs with SOFI, the infrared
spectro-imager at the NTT (3.5-m telescope, la Silla, ESO). 
Our goal is to identify faint companions over the full accessible
separation range, from the 0.5-0.8$"$ typical infrared seeing at La Silla to
the $\sim 5'$ field of view of the images. We therefore obtain multi-epoch 
observations to separate true companion from background stars through 
common proper motion. The companion of DENIS-J055$-$44 however
is sufficiently bright that the DENIS images could provide the 
first epoch. We therefore marked the pair for early follow-up.

We use SOFI in its Large Field mode, which provides
0$"$.288/pixel and a 4.9$'$${\times}$4.9$'$ field of view. The
observation sequence consists of 5 individual 1~mn exposures acquired on 5
offset pattern within a few arcsecond jitter box. The reduction uses
the ECLIPSE package (Devillard et al. \cite{dev97}). The
individual raw data are flat-fielded using a normalized gain map, derived
from images of the illuminated dome. The sky signal is estimated from a
median across the jittered images and subtracted from the flat-fielded
image. Finally, the 5 flat-fielded and sky-subtracted images of each source
are shifted and averaged to produce the final image. With nominal weather
conditions the 5$\sigma$ limiting magnitude is $J$=21.5. 

\begin{figure}
\includegraphics[width=9cm,angle=-90]{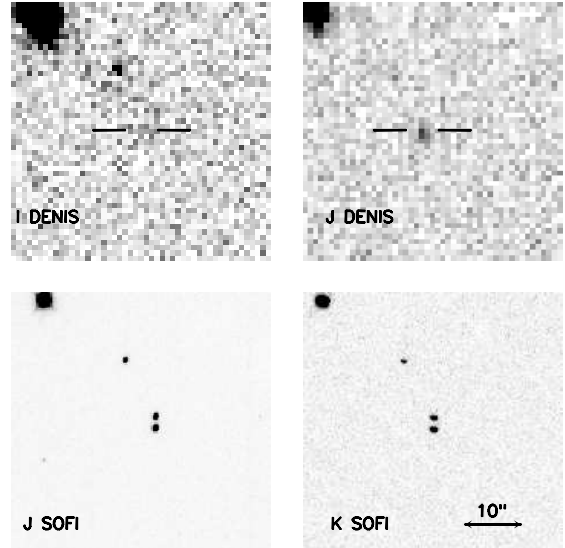}
\caption{DENIS discovery images in $I$ and $J$ (top), and SOFI images
demonstrating the binary status in $J$ and $K$ (bottom). North is up and East
is left. 
}
\label{image}
\end{figure}

The NTT images (Fig.~\ref{image}) show a well separated double star with 
a moderate contrast of 0.5~mag and a separation of $\sim~2.2"$ 
(Table~\ref{table_phot}). The shallower and lower resolution DENIS
images show a single source that is marginally elongated at $J$ and
only barely detected at $I$, with joint magnitudes of $J=15.3\pm0.15$ 
and $I=18.4\pm0.3$. DENIS did not detect the target at $K_s$. The lack 
of any bright source within 15$''$ of DENIS-J055$-$44 in the DENIS $I$ image 
demonstrates that the second component is not a blue background star: 
even assigning the whole $I$ flux to the fainter component would still 
leave it with the colour of at most a mid-M dwarf. At the DENIS magnitude 
limit the sky density of mid-M dwarfs is below 1 per square degree, and 
the probability to find one by chance within 2.2$''$ of one of only 250 
targets is thus negligible. DENIS-J055$-$44 also figures in the 2MASS 
catalog (Cutri et al. \cite{cutri03}) as a single source, with the $J-K$ 
colour of a late-M dwarf (Table~\ref{table_phot}). 

\section{Infrared spectroscopy}
The spectra were also obtained with SOFI, on the nights of 
February 9$^{th}$ and 10$^{th}$ 2003. We used the blue and red 
low-resolution  grisms, for dispersions of 6.96 and 
10.22 \AA/pix, wavelength ranges of respectively 0.95-1.64 and 
1.53-2.52~${\mu}m$, and a resolution of 600 with the 1$\arcsec$ slit.
SOFI was rotated to align the two components on the slit, and we used
the standard observing template for SOFI spectroscopy: the objects are 
nodded between two positions along the slit, with a
small jitter to fill in the bad pixels. The total exposure time is 9
min for the blue part of the spectrum and 6 min for the red part. 
A standard star (HIP 30374, F5V) was observed immediately after the target 
for flux calibration.

\begin{figure*}
\begin{center}
\includegraphics[width=7cm]{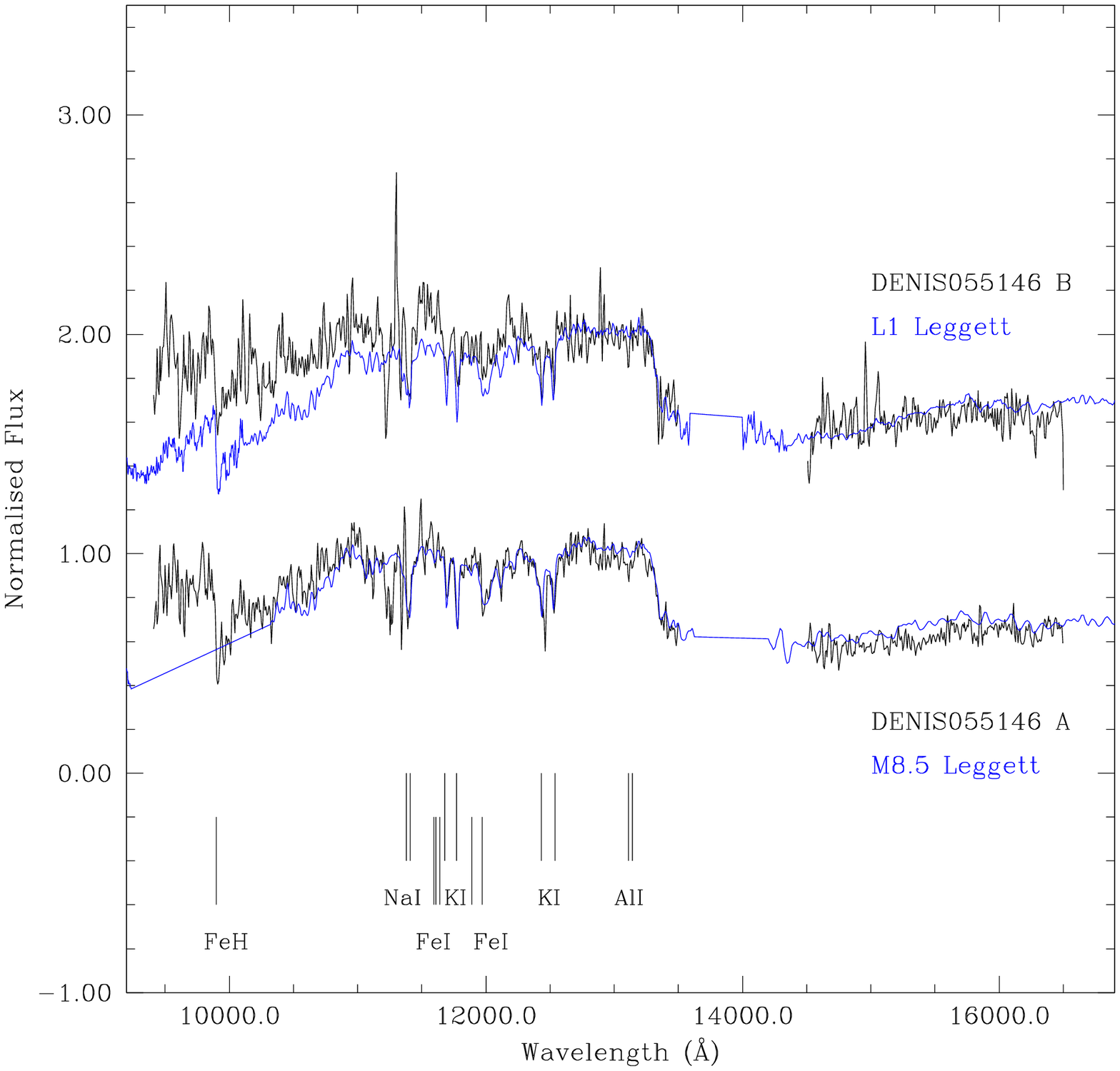}
\includegraphics[width=7cm]{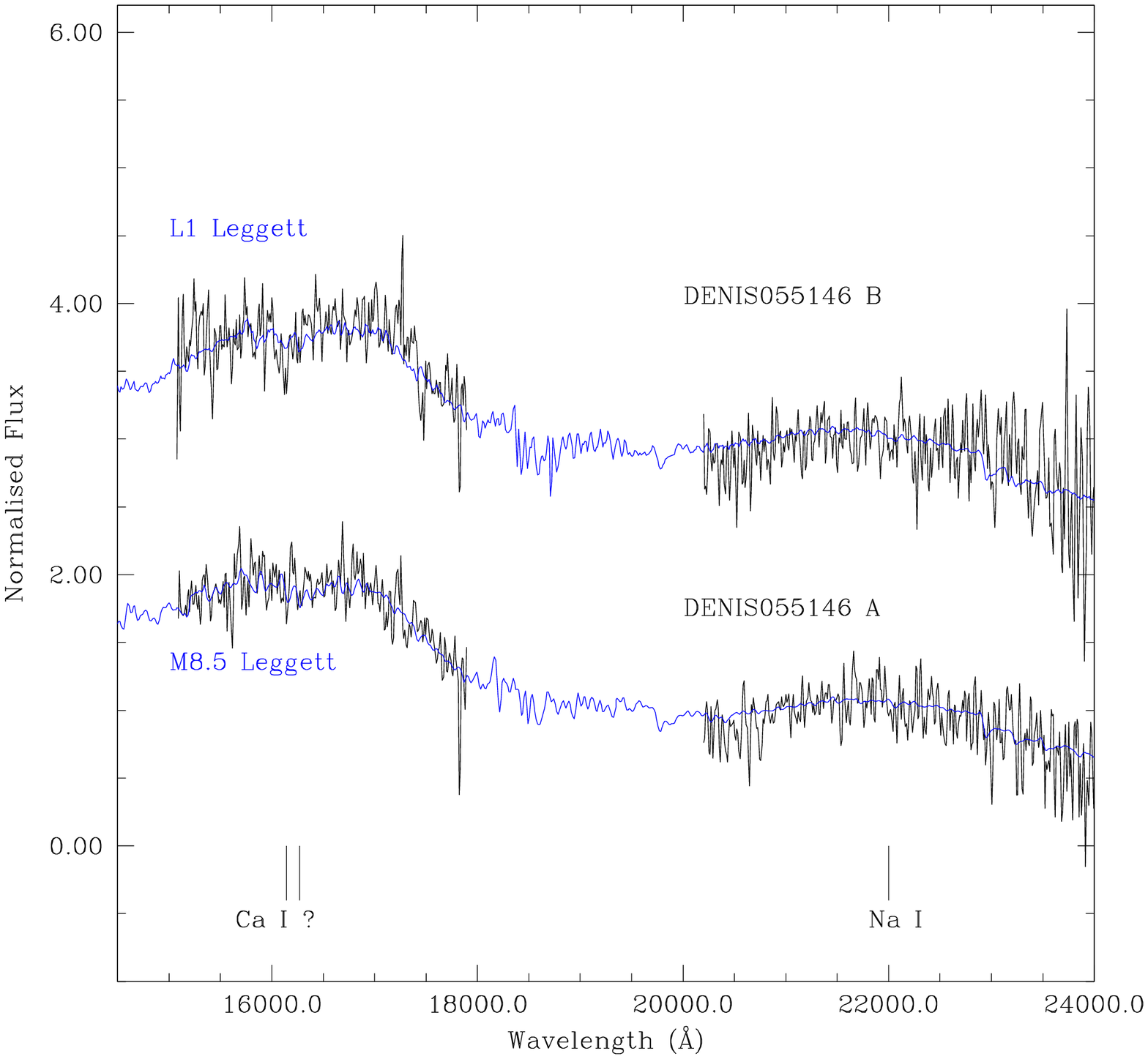}
\caption{Blue (left panel) and red (right panel) spectra of the two 
components of DENIS~0551$-$44. The blue and red spectra are 
normalized at respectively 12650~\AA ~and 20800~\AA, and vertically 
shifted for clarity. The blue overlays represent spectra of Leggett
spectral templates, with spectral types M8.5 for the primary and
L1 for the secondary (the Leggett database does not currently contain 
an infrared L0 spectrum). The integration ranges of the spectral indices
are marked at the top, and several characteristic spectral features
at the bottom. }
\label{denis055_spectra}
\end{center}
\end{figure*}

The spectral images were flat-fielded, sky-subtracted and recombined 
with a combination of a MIDAS script and ECLIPSE procedures (isaacp 
package). The images were then rectified, and the spectra of the two
components were extracted and wavelength calibrated with IRAF tasks. 
The wavelength calibrations is based on spectra of a Xenon lamp
obtained on the same night, and is accurate to better than 1 \AA. 
To correct for the instrumental spectral response and for telluric 
absorption, the spectra were then divided by the ratio of the
spectrum of the spectrophotometric standard to a Kurucz model of
an F5V star (Kuruzc \cite{kuru93}). Due to the modest altitude of La Silla
observatory and the summer observing, the strong telluric water
absorption bands are imperfectly corrected. Figure~\ref{denis055_spectra} 
therefore blanks out their wavelength range. The wider stellar
$H_2O$ bands remain conspicuous and demonstrate the late-M/early-L
status of the two stars.

The spectra have moderate resolution and relatively poor S/N ratios, in 
particular for the secondary. Our attempts at using spectral indices 
(Delfosse et al. \cite{delf99}; McLean et al. \cite{mclean03}) to derive quantitative 
spectral types were therefore only partly successful. We instead use
the full available spectral range, and determine spectral types through
least square-matching to the spectral library made available by S.~Leggett
(http://www.jach.hawaii.edu/pub/ukirt/skl/dM.spectra/). In the least square
matching we ignore the telluric H$_2$O bands (1.34-1.43$\mu$m and
1.80-2.00$\mu$m), as well as the $\lambda < 1.1 \mu$ m range which
is missing for some of the Leggett templates. Both parts of the
primary spectrum give a well defined ${\chi}^2$ minimum at spectral 
type M8.5, while the noisier spectrum of the secondary gives acceptable 
matches for spectral types M9 to L1. Figure~\ref{denis055_spectra} overlays 
the spectra of the two components with respectively an M8.5 and an L1 
template. The M8.5 spectrum clearly matches DENIS~0551$-$44A very well. The L1 
spectrum provides a fair match to DENIS~0551$-$44B, as does an M9
template, but it leaves some residuals between 1 and 1.2$\mu$m. This 
could quite possibly be due to some contamination from light of the 
close primary, and better quality spectroscopy would be desirable.
In the mean time we adopt a spectral type of L0$\pm$1 for DENIS~0551$-$44B.
The photometric contrast of $\Delta~J$=$\Delta~K$=0.5 corresponds
to a difference of 1.5 spectral sub-type for late-M or early-L dwarfs 
(Dahn et al. \cite{dahn02}), and strengthens the adopted classification
of the secondary.

The sky density of $I-J \ge 3.0$ (later than M8) objects above
the DENIS detection limit ($I = 18.5$) is only 0.052/square degree
(Delfosse et al. \cite{delf03}). Each component of DENIS~0551$-$44 being 
individually fainter than this limit, we conservatively increase the 
detection volume by $2^{1.5}$. Adopting a constant stellar density 
(again a slightly conservative assumption), 
this increases the density of ultracool dwarfs brighter than 
DENIS~0551$-$44B to ${\lesssim}$ 0.15/square degree. 
The probability of finding two unrelated late-M or L dwarfs 
with a projected separation below 5$''$ is thus ${\lesssim} 3.10^{-7}$.
Moreover, the magnitude difference of the two components is consistent 
with their spectral type, indicating identical distances to within $\sim20\%$. 
We will soon obtain second-epoch IR images to demonstrate the common proper 
motion of the pair, but their physical association is already beyond 
reasonable doubt.

Correcting the 2MASS $J$ magnitude of the system for the companion results
in a primary magnitude of $J_A=16.3$. Together with the Dahn et 
al. (\cite{dahn02}) spectral type versus M$_{\rm J}$ relation 
(M$_{\rm J} \sim 11.25$ for M8.5), this results in a spectroscopic 
distance of $\sim~100$~pc for the system and a projected separation 
between the two components of $\sim$220~AU.

\section{Discussion}
Numerous previous studies (Close et al. \cite{close03}; Burgasser et
al. \cite{burg03}; Bouy et al. \cite{bouy03}; Gizis et al. \cite{gizis03};
Siegler et al. \cite{sieg05}) had found no wide binary ultracool dwarfs 
in the field, and argued that such systems either never form or must be
quickly disrupted during the first few Myr of their life. DENIS~0551$-$44 
demonstrates that wide binary ultracool dwarfs do exist in the field,
as they were known to at earlier stages of their evolution (Luhman
et al. \cite{luh2004}; Chauvin et al. \cite{chau04}, \cite{chau05}).
Wide ultracool binaries are undoubtedly rare in the field, and the lack 
of any detection in previous studies most likely reflects
no more than their smaller samples. 

These wide binaries are unexpected in the embryo-ejection scenario, 
probably the leading theory for the formation of ultracool dwarfs. In 
current hydrodynamical simulations of the fragmentation of small molecular 
clouds (Delgado-Donate et al. \cite{delg04}; Bate et al. \cite{bate03}), 
ultracool dwarfs only form in unstable multiple systems, from which they 
are ejected before they can accrete most of the mass available in their
reservoir. These strong early dynamical interactions obviously disrupt 
any wide binary, because of its weak gravitational bound.

One major motivation behind the embryo-ejection scenario is that 
the typical thermal Jeans mass in molecular cloud cores is 
$\sim$1M$_{\odot}$ (Larson \cite{larson99}), so classical
gravitational collapse only forms stars and no brown dwarfs. 
Padoan \& Nordlund (\cite{pado04}), however, show that accounting
for turbulence allows gravitational collapse to form brown
dwarfs as well: supersonic turbulence generates highly nonlinear 
density and velocity field in star-forming clouds, invalidating 
the quasi-static Jeans gravitational stability criterion. Magnetic 
fields in addition most likely play an important role in the 
fragmentation/collapse process and their effects on the IMF 
are poorly understood. Brown dwarfs (and object at the stellar-substellar 
boundary) could thus well form like stars, in which case the small
number of wide binaries rather reflects a posterior stage of dynamical 
decay for most weakly bound systems (Sterzik \& Durisen \cite{sterzik03}).

An alternative formation process for ultracool dwarfs is the 
photo-evaporation model (Kroupa \& Bouvier \cite{kroupa03}; Whitworth \&
Zinnecker \cite{Whit04}), where very low mass objects form in the vicinity
of OB stars whose strong UV fields erodes the outer layers of protostellar
cores before reach a stellar mass. This mechanism is only effective in 
(very) dense star formation regions, and highly wasteful of the available 
gas. It could thus be responsible for a few very low mass objects, but not 
for a majority.

DENIS~0551$-$44 clearly did not form through embryo-ejection, but could
be the result of either fragmentation, like more massive stars, or 
of photo-evaporation in the vicinity of OB stars. As previously
discussed by Luhman (\cite{luh2004}), the existence of a few very 
wide ultracool binaries demonstrates that some brown dwarfs form
outside the embryo-ejection scenario, though not that all of them
do. \footnote{A proviso to this statement is that DENIS~0551$-$44 could in 
principle be of higher multiplicity, and hence have a larger
mass than is apparent. Higher resolution imaging would be of
interest to ascertain this.} The proper-motion analysis of our 
full sample will shortly provide a good measurement of the 
fraction of such wide binaries among ultracool dwarfs. This 
statistics will constrain the relative contributions of the 
various modes of brown dwarf formation.

\begin{acknowledgements}

We thanks the NTT team for their precious help during the observations 
which led to these results. We are grateful to the referee, K.L. Luhman
for suggestions that improved the paper and a prompt report. Financial 
support from the "Programme National de Physique Stellaire'' (PNPS) 
of CNRS/INSU, France, is gratefully acknowledged.
DENIS is the result of a joint effort involving human and financial
contributions of several Institutes mostly located in Europe. It has
been supported financially mainly by the French INSU, CNRS, and French 
Education Ministry, the ESO, the State of Baden-Wuerttemberg, and
the European Commission under networks of the SCIENCE and Human
Capital and Mobility programs, the Landessternwarte, Heidelberg,
l'Institut d'Astrophysique de Paris, the Institut fur Astrophysik der
Universitat Innsbruck and Instituto de Astrofisica de Canarias.

\end{acknowledgements}

\end{document}